\documentclass[a4paper,11pt]{article}
\usepackage{pos}

\title{Composite nature of exotic states from data analysis}

\author*[a]{Xian-Wei Kang}

\affiliation[a]{Key Laboratory of Beam Technology of Ministry of Education,
School of Physics and Astronomy,\\ Beijing Normal University, Beijing 100875, China.}

\emailAdd{xwkang@bnu.edu.cn}

\abstract{We introduce two basic concepts: the compositeness for resonance and the Castillejo-Dalitz-Dyson 
(CDD) pole. Applying them to hadron states $X(3872)$,  $Z_b(10610)$ and $Z_b(10650)$, and $T_{cc}$, we obtain their compositeness values and achieve deeper insights into their inner structure.}

\FullConference{The 21st International Conference on Hadron Spectroscopy and Structure (HADRON2025)\\
 27 - 31 March, 2025\\
Osaka University, Japan\\}

\begin{document}
\maketitle

\section{Introduction}
Since the discovery of $X(3872)$ by Belle collaboration in 2003 \cite{Belle:2003nnu}, a series of exotic hadron candidates were and are observed. They cannot be accommodated by potential model. So the interpretations, such as the kinematic effect, molecule, quark-gluon hybrid, are proposed. However, typically different scenarios predict different constituents. And in practice, several mechanisms may involve. The concept of the compositeness will then be important and needed. At the mean time, a notable feature is that such states often stay close to threshold of the hadron pair, e.g., only 2-3 MeV above the threshold. The effective range expansion (ERE) works in this region for treating the two-body scattering. In our present approach, the scattering amplitude with inclusion of CDD poles applies more widely than ERE method. 

\section{Compositeness for resonance}
The compositeness $X$ gives the weight of two-hadron componenets in the configuration of a molecule state. $X=0$ indicates an elementary state while $X=1$ indicates a pure molecule. The concept of the compositeness is dated from Weinberg's definition \cite{Weinberg:1962hj,Weinberg:1965zz}. For a pure molecule, one has its wave function renomalization constant $Z=0$. The compositeness $X=1-Z$, with $X$ defined as $X=-\gamma^2\frac{dG(s_R)}{ds_R}$ quantifying the weight of its constituents. $\gamma^2$ is the residue of the scattering matrix $t(s)$ in the physical Riemann sheet at the pole $s_R$, and $G(s)$ is the two-point loop function. This definition is only applicable to bound state, and based on it, the model-independent relation between scattering length, effective range and compositeness $X$ is given for the deuteron. 

For a resonance case, the above mentioned $X$ becomes complex-valued and then loses the physical interpretation of the weight. We will adopt the generation made by Guo and Oller in Ref.~\cite{Guo:2015daa}. As long as $\sqrt{\text{Re}s_R}$ is larger than the lightest threshold (for coupled channel), $X=|\gamma^2\frac{dG(s_R)}{ds_R}|$, with $\gamma^2$ the residue of $t(s)$ in the unphysical Riemann sheet where the resonance appears in. This criterion is adapted to the non-relativistic case: $M_R>m_\text{th}$ where $M_R$ is the resonance mass and 
$m_\text{th}$ the mass threshold of the hadron pair \cite{Kang:2016ezb}. For other definitions of compositeness, see also Refs.~\cite{Wang:2022vga,Gao:2018jhk}, where many interesting findings are found in the study of $f_0(980)$ and $a_0(980)$. 

\section{CDD pole and the scattering amplitude}
The introduction of the CDD pole orginates from the study of Low's equation \cite{Castillejo:1955ed}. There it is mentioned that the scattering amplitude can contain infinite number of adjustable parameters, and nowadays they are the CDD poles and corresponding residues that cannot determined by the scattering amplitude itself. 

The scattering amplitude with inclusion of only the right-hand cut can be written by
\begin{eqnarray}
t(s)=\left[\frac{\gamma^2}{s-M_{\text{CDD}}^2}+G(s)\right]^{-1},
\label{eq:ts-rela}
\end{eqnarray}
where we have considered only one possible CDD pole ($M_{\text{CDD}}$ and its residue $\gamma^2$). $G(s)$ is the two-point loop function which can be written as 
\begin{equation}
G(s)=\alpha(\mu^2)+\frac{1}{(4\pi)^2}\left(\log\frac{m_2^2}{\mu^2}-\varkappa_+\log\frac{\varkappa_+-1}{\varkappa_+}
-\varkappa_-\log\frac{\varkappa_--1}{\varkappa_-} \right),
\end{equation}
with
\begin{equation}
\begin{split}
\varkappa_{\pm}&=\frac{s+m_1^2-m_2^2}{2s}\pm\frac{k}{\sqrt{s}}\,,\\
k&=\frac{\sqrt{(s-(m_1-m_2)^2)(s-(m_1+m_2)^2)}}{2\sqrt{s}}\,,
\end{split}
\end{equation}
in the dimensional regularization form. $\alpha(\mu^2)$ being the subtraction constant at the renormalization scale $\mu$. Changing the scattered particle mass $m_2$ by $m_1$ 
does not matter for the final conclusion. Equation \eqref{eq:ts-rela} can be obtained from the N/D method \cite{Oller:1998zr}. For the recent pedagogical material, one could refer to Refs.~\cite{Oller:2024lrk,Oller:2025leg}. 

We remark on the CDD poles:
\begin{itemize}
\item Inclusion of CDD poles does not violate constraints from analyticity and unitarity.
\item CDD pole is the zero of $t$-matrix amplitude. 
\item The location of CDD pole is related to the compositenss $X$. That is, if $M_\text{CDD}$ is far from the threshold, the two-hadron molecule dominates; if $M_\text{CDD}$ is close to the threshold, the state is more ``elementary'' with small value of $X$.
\end{itemize}
In the non-relativistic limit, we have
\begin{equation}\label{eq:tE-nonrela}
t(E)=8\pi m_\text{th}\left[\frac{\lambda}{E-M_{\text{CDD}}}+\beta-ik\right]^{-1},
\end{equation}
with
\begin{eqnarray}
\lambda&=&\gamma^2\frac{8\pi m_\text{th}}{m_\text{th}+M_\text{CDD}},\nonumber\\
\beta&=&8\pi m_\text{th}\alpha(\mu^2)+\frac{1}{\pi}\left(m_1\log\frac{m_1}{\mu}+m_2\log\frac{m_2}{\mu}\right).
\end{eqnarray}
Comparing to the effective range expansion formalism:
\begin{equation}
t(E)=8\pi m_\text{th}\left(-\frac{1}{a}+\frac{1}{2}r k^2 -i k\right)^{-1},
\end{equation}
we have the scattering length $a$ and effective range $r$ as 
\begin{equation}
\begin{aligned}
\frac{1}{a}&=-\frac{\lambda}{m_\text{th}-M_{\text{CDD}}}-\beta,\\
r&=-\frac{\lambda}{\mu (m_\text{th}-M_{\text{CDD}})^2}.
\end{aligned}
\label{eqar}
\end{equation}
A prominent feature of the above equation is that when the CDD pole stays very close to threshold, the effective range $r$ will become huge, which raise doubts on the validity of effective range expansion. In such case, our present parameterization does work.  

\section{Application to hadron states $X(3872)$, $Z_b(10610)$ and $Z_b(10650)$, and $T_{cc}$}
The final state interaction is meadited by the $d(E)$ function as
\begin{equation}\label{eq:dEgeneral}
d (E) = \left(1 + \frac{E - M_{\text{CDD}}}{\lambda} (\beta - \mathrm{i} k)\right)^{-1},
\end{equation}
which differs from $t(E)$ by removing the extra zero at CDD pole. $|d(E)|^2$ (not $|t(E)|^2$ ) provides the parameterization of the signal shape. One can compare it to a standard Breit-Wigner shape, referring to Fig.~2 in Ref.~\cite{Guo:2016wpy}. Considering the experimental resolution effect, finite bin width, background shape, one could write out the energy-dependent event distribution (spectrum). By fitting to experimental data, one can fix the parameter. Once the full amplitude is known, we can search for the pole in the complex plane \cite{Kang:2016zmv}, calculate its residue corresponding to the pole, and calculate the compositeness value defined above as well.   

We have made a careful analysis of $X(3872)$ data, and the results were given in Refs.~\cite{Kang:2016jxw,Kang:2019tgv}.
The experimental data is reproduced rather well in several scenarios. $X(3872)$ can be a bound and/or a virtual state, and also might be a higher-order virtual-state pole (double or triplet pole) in the limit of vanishing $D^{*0}$ width. The compositeness could range from 0 to 1. Recently, we also analyze the new BESIII data \cite{BESIII:2023hml} using the scattering length approximation \cite{Kang:2024gxg} and find that $X(3872)$ could be understood as a molecule of $D\bar D^*$. Note that the BESIII data \cite{BESIII:2023hml} refers to the spectrum data on $D^0\bar D^0\pi^0$, different from the analysis in
Ref.~\cite{Kang:2016jxw} with the spectrum data on $D^0\bar D^{*0}$. The radiative decays $X(3872)\to J/\psi\gamma$ and $\psi(3686)\gamma$ are also studied in the covariant light-front quark model. It shows that there is little chance for $X(3872)$ to be understood as a pure charmonium \cite{Yu:2023nxk}. 

\begin{figure}[ht]
\begin{minipage}{1.0\linewidth}
\centerline{\includegraphics[width=0.5\textwidth]{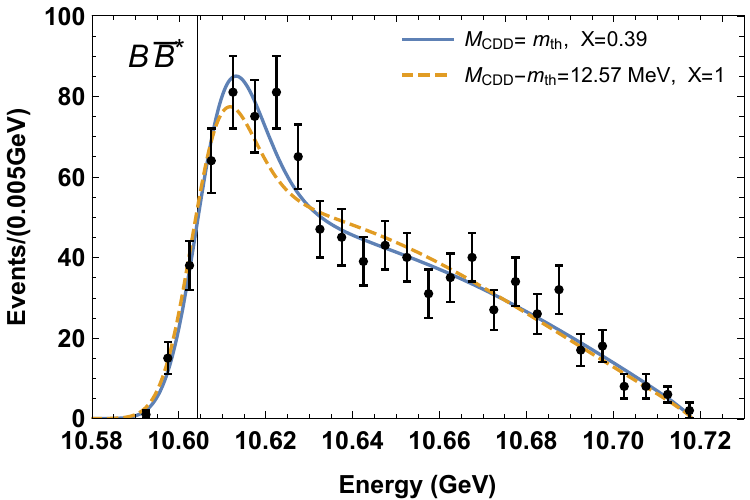}\;\;\;\includegraphics[width=0.5\textwidth]{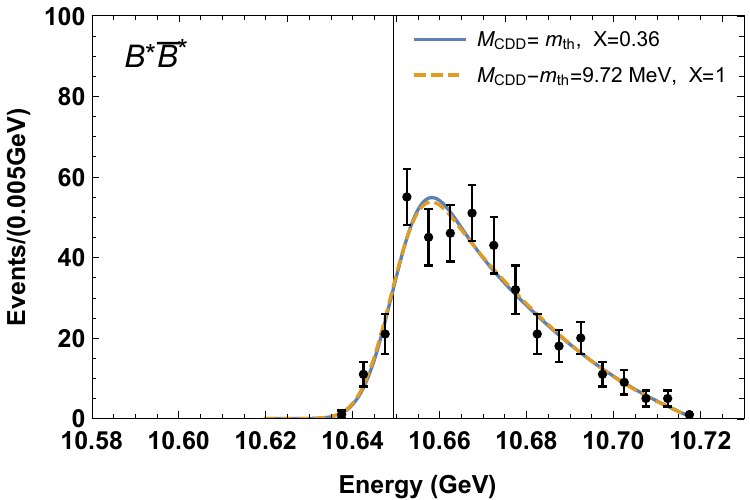}}
\end{minipage}
\caption{Mass spectra of $Z_b(10610)$ and $Z_b(10650)$. The left (right) one is for $B\bar B^*$ ($B^*\bar B^*$) invariant mass spectrum and the relevant $Z_b(10610)$ ($Z_b(10650)$) resonance. Points with error bar are the experimental data from Belle Collaboration \cite{Belle:2015upu}. The vertical lines correspond to the $B^{(*)}\bar{B}^*$ thresholds. In each subfigure, the solid line corresponds to $M_{\text{CDD}}=m_\text{th}$ case, where the value of compositeness $X$ is smallest, 0.39 for $Z_b(10610)$ and 0.36 for $Z_b(10650)$; the dashed line corresponds to $X=1$. }
\label{figvs}
\end{figure}

Similarly, assuming $Z_b(10610)$ and $Z_b(10650)$ as resonances of $B\bar B^*$ and $B^*\bar B^*$, respectively, the data can be well reproduced, as shown in Fig.~\ref{figvs} \cite{Zhang:2022hfa}. The parameters of $M_{\text{CDD}}$ can vary in a very wide range such that
the compositenesses of both $Z_b$ take the values of $0.4\sim 1$.

\begin{figure}[htbp]
    \centering
    \includegraphics[width=0.8\textwidth]{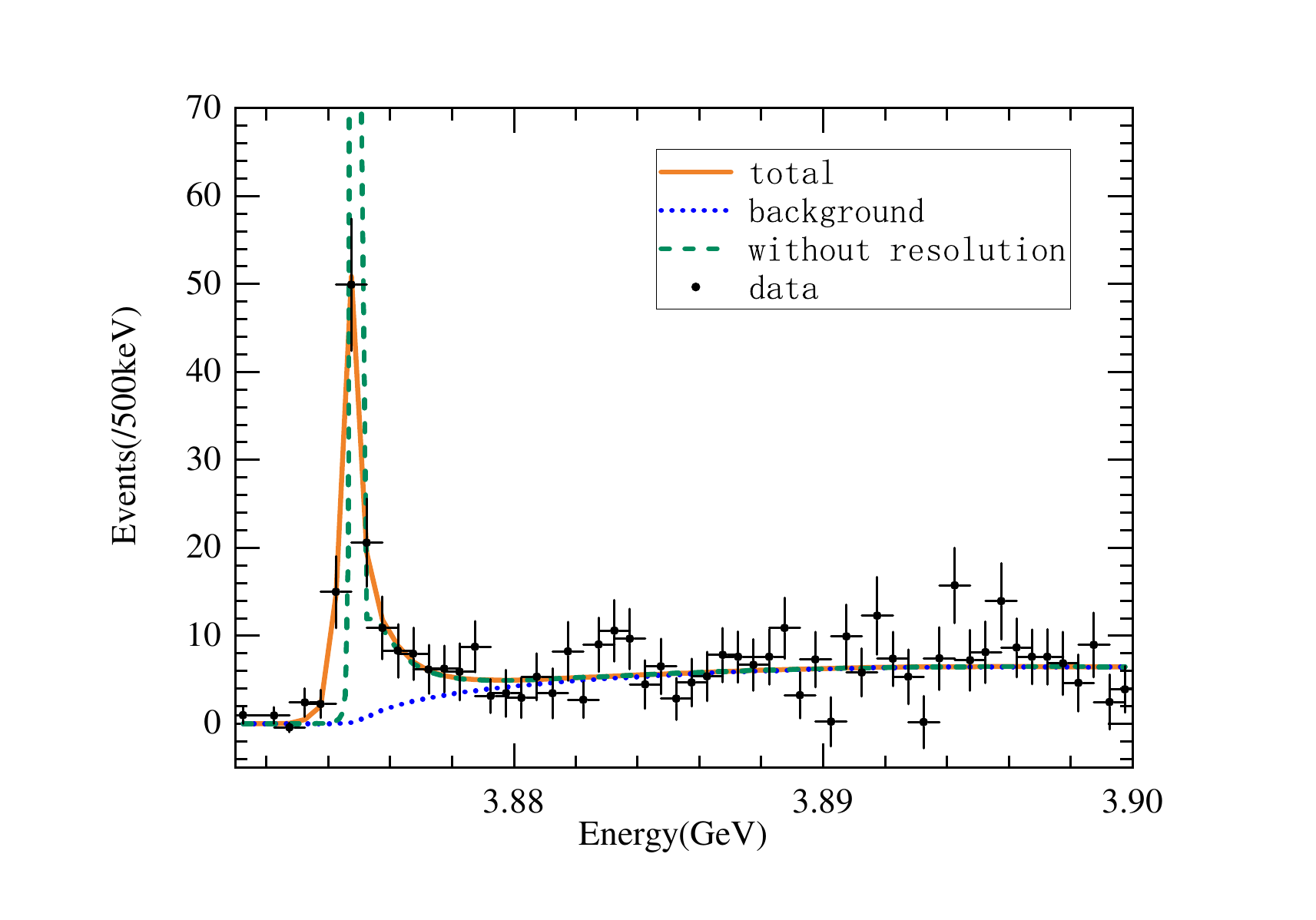}
    \caption{Mass spectrum for the $D^0D^0\pi^+$ decay channel. The data are from the LHCb collaboration \cite{LHCb:2021auc}. The solid line represents our total result, with dotted line showing the background and dashed line corresponding to without the Gaussian resolution.}\label{fig:fig7n}
\end{figure}

Very recently, we have also analyzed the $T_{cc}$ data \cite{Kang:2024jvt}. With the fitted parameters, we have the spectrum result in Fig.~\ref{fig:fig7n}. Interestingly, both poles are found in the physical and unphysical Riemann sheets, which indicates that $T_{cc}$ has large portion of "elementary"
degree of freedom, e.g., the tetraquark, by considering the Morgan rule \cite{Morgan:1992ge}. The compositeness as a measure of molecule component in its wave function
is predicted to be $0.23^{+0.40}_{-0.09}$. A detailed and more rigorous study in the coupled channel is ongoing.

\section{Conclusion and outlook}
We proposed a new parameterization with inclusion of the CDD pole to treat the nontrivial and nonperturbative two-body interaction. By fitting to the experimental data, we can fix the appearing parameters. Then, the hadron property can be examined and the compositeness can be quantatified if the hadron could be interpretated as a molecule state. Successful applications to $X(3872)$, $Z_b(10610)$ and $Z_b(10650)$, and $T_{cc}$ have been done and introduced in this Proceeding. 

\acknowledgments
 The author is indebted to J.~A.~Oller and Zhi-Hui Guo for fruitful discussion and collaboration in some of the publications. This work is supported by the National Natural Science Foundation of China (NSFC) under Project No. 12275023.

\end{document}